# Extending the FDTD GVADE method nonlinear polarization vector to include anisotropy

**CALEB J. GRIMMS**[*] **AND ROBERT D. NEVELS**

*Department of Electrical and Computer Engineering, Texas A&M University, 188 Bizzel St, College Station, TX 77801, USA*
**cgrimms@tamu.org*

**Abstract:** In this paper the finite-difference time-domain general vector auxiliary differential equation method [Greene, J. H. and A. Taflove, Opt. Express **14**, 8305 (2006)], nonlinear polarization vector, the nonlinear electric dipole moment per unit volume, is extended to include anisotropy, in nonlinear isotropic media at optical frequencies. The theory is presented for extending the numerical method in 3D Cartesian coordinates, and then example simulation results are presented for two isotropic media. First, the simplified 2D transverse magnetic case is revisited for the fused silica example introduced in the 2006 paper, including the anisotropic part of the nonlinear polarization vector in the simulation; the simulation results including the anisotropic part of the polarization vector were compared with the purely isotropic polarization vector simulation results. Second, a simplified 2D transverse magnetic example was simulated in carbon disulfide, with its strong molecular re-orientation induced polarization anisotropy.

## 1. Introduction

In this paper, the mathematics for the simulation of electromagnetic wave propagation in nonlinear isotropic material at optical frequencies using the general vector auxiliary differential equation finite-difference time-domain (GVADE FDTD) numerical method [1-3] are discussed, extending and generalizing the method's polarization vector to include an anisotropic part. The FDTD GVADE method models nonlinear behavior in nonlinear isotropic material using polarization current density from Ampere's law, which is related to the polarization vector by a time-derivative. The method assumes the nonlinear material is isotropic, meaning that without any electric or magnetic field applied the average material electromagnetic properties at a macroscopic level are the same in any direction, in contrast to many crystals for example where the material electromagnetic properties vary based on the direction. The FDTD GVADE method also assumes the polarization vectors in Maxwell's equations are isotropic, meaning "the polarization is assumed to line in the direction of the electric field" according to page 500 of [3] and the average material response, the electric flux density, is equally proportional to the applied electric field in any direction at a given point. However, electromagnetic wave propagation in some isotropic media can result in a polarization vector with an anisotropic part along with the isotropic part. This anisotropic part means the average material response, electric flux density, from an applied electric field is not the same in every direction, requiring a tensor to describe properly. The purpose of this paper is to extend the FDTD GVADE method to simulate electromagnetic wave propagation in nonlinear isotropic material where both the isotropic part and the anisotropic part of the polarization vectors are present.

Historically, many authors have worked on expanding the FDTD method [4, 5] to model electromagnetic waves in nonlinear media, such as [6-13] for a few examples. Authors have also worked on expanding the FDTD method to model electromagnetic wave propagation in anisotropic media as well with [14-19] as some examples. This paper is not simulating electromagnetic wave propagation in anisotropic media. The goal of this paper is attempting to accurately model electromagnetic wave propagation in a nonlinear isotropic media with a polarization vector including an anisotropic part.

As a brief, non-comprehensive historical summary, optical frequency intensity dependent "self-induced" polarization changes have been observed since the 1960s [20-24], allowing phenomena like the "self-focusing" of electromagnetic waves to occur. During the 1960s and 1970s, the theory behind these phenomena began to be developed [25-31], and can be described and incorporated into Maxwell's equations using the nonlinear polarization vector. The classic review paper, discussing polarization vector at optical frequencies is R. W. Hellwarth's 1977 paper [28]. Along with the general discussion of the polarization density vector in macroscopic material, Hellwarth's 1977 paper also presents a simplification for the polarization vector at optical frequencies in liquids and solids, the Born-Oppenheimer approximation, for isotropic media and anisotropic media like crystals with different types of symmetries such as cubic, hexagonal, trigonal, etc. In this paper we are only considering isotropic media, and as such, will limit our discussion to the Born-Oppenheimer approximation of the polarization vector for isotropic media.

## 2. Theory

The advantage of FDTD based methods [4, 5] such as the FDTD GVADE method [1-3], is that they are generally considered to be more accurate than many other methods used in optics, such as the Nonlinear Schrödinger (NLS) equation. The potential increased accuracy is due to FDTD based methods needing fewer assumptions and simplifications than many other methods, solving the full vector Maxwell's equations directly, while requiring more computational resources as a result [32].

*2.1 Maxwell's equations in isotropic media using vector notation*

Beginning by writing Maxwell's equations for a nonmagnetic material without any free charges $\rho$ or impressed current density vector **J**:

$$\nabla \times \mathbf{H} = \frac{\partial \mathbf{D}}{\partial t}, \tag{1}$$

$$\nabla \times \mathbf{E} = -\mu_0 \frac{\partial \mathbf{H}}{\partial t}, \tag{2}$$

where **H** is the instantaneous magnetic field vector and **D** is the electric flux density. The Ampère-Maxwell law can be re-written in terms of the instantaneous electric field vector **E**, the induced polarization current density vector **J** and induced polarization density vector **P** defined as the "electric dipole moment per unit volume", where $\mathbf{D} = \varepsilon_0 \mathbf{E} + \mathbf{P}$ and $\mathbf{J} = \frac{\partial \mathbf{P}}{\partial t}$ [3, 33, 34]:

$$\nabla \times \mathbf{H} = \varepsilon_0 \frac{\partial \mathbf{E}}{\partial t} + \mathbf{J}, \tag{3}$$

The polarization current density vector **J** can be re-written by expanding the polarization vector **P** into a series expansion of the electric field **E**, where $\mathbf{P} = \mathbf{P}^{(1)} + \mathbf{P}^{(2)} + \mathbf{P}^{(3)} + \cdots$. For many optical materials, the $\mathbf{P}^{(2)}$ term is not present [28, 34]. The equations for $\mathbf{P}^{(1)}$ and $\mathbf{P}^{(3)}$ are

$$\mathbf{P}^{(1)}(\mathbf{r},t) = \varepsilon_0 \int_{-\infty}^{\infty} [\chi^{(1)}(t-t') \cdot \mathbf{E}(\mathbf{r},t')] dt', \tag{4}$$

$$\mathbf{P}^{(3)}(\mathbf{r},t) = \varepsilon_0 \int_{-\infty}^{\infty} \int_{-\infty}^{\infty} \int_{-\infty}^{\infty} [\chi^{(3)}(t-t_1, t-t_2, t-t_3) \vdots \mathbf{E}(\mathbf{r},t_1)\mathbf{E}(\mathbf{r},t_2)\mathbf{E}(\mathbf{r},t_3)] dt_1 dt_2 dt_3, \tag{5}$$

where the tensor products of electric field with the susceptibility tensors $\chi$ are shown [3, 34] using the "·" and ":" notation. The "first-order" term $\mathbf{P}^{(1)}$ is called "linear" due to $\mathbf{P}^{(1)} \propto E$ while the "third order" term $\mathbf{P}^{(3)}$ or $\mathbf{P}^{NL}$ is called "nonlinear" due to $\mathbf{P}^{(3)} \propto (E)^3$. The linear $\mathbf{P}^{(1)}$ term used here is the same as the FDTD GVADE method [1-3], modelling linear Lorentz dispersion using a Sellmeier expansion, and as a result, it will not be discussed here. The first-order susceptibility $\chi^{(1)}$ is a rank 2 tensor containing 9 terms, and it is assumed to be isotropic here. The third-order susceptibility $\chi^{(3)}$ is a rank 4 tensor containing 81 terms, which is also assumed to be isotropic, but it can lead to a polarization vector with an anisotropic part. For many materials at optical frequencies, the Born-Oppenheimer Approximation can be used to simplify $\mathbf{P}^{(3)}$ [28], separating out the much faster "electronic" response from the slower "nuclear" response, assuming the "electronic" response is effectively instantaneous in comparison,

$$\mathbf{P}^{NL}(\mathbf{r},t) = \mathbf{P}^{NL}_{el}(\mathbf{r},t) + \mathbf{P}^{NL}_{nu}(\mathbf{r},t), \tag{6}$$

$$\mathbf{P}^{NL}_{el}(\mathbf{r},t) = \varepsilon_0 \chi^{(3)}_{el} \vdots [\mathbf{E}(\mathbf{r},t)\mathbf{E}(\mathbf{r},t)\mathbf{E}(\mathbf{r},t)] = \varepsilon_0 \chi^{(3)}_{el} \mathbf{E}(\mathbf{r},t)|\mathbf{E}(\mathbf{r},t)|^2, \tag{7}$$

$$\mathbf{P}^{NL}_{nu}(\mathbf{r},t) = \varepsilon_0 \mathbf{E}(\mathbf{r},t) \int_{-\infty}^{\infty} dt_1 [\chi^{(3)}_{nu}(t-t') : \mathbf{E}(\mathbf{r},t')\mathbf{E}(\mathbf{r},t')]. \tag{8}$$

For an isotropic media, this equation can be re-written as [26, 28, 34]

$$\mathbf{P}^{NL}_{nu} = \varepsilon_0 \mathbf{E}(\mathbf{r},t) \int_{-\infty}^{\infty} \left[ \chi^{(3)}_{nu,a}(t-t') |\mathbf{E}(\mathbf{r},t')|^2 \right] dt' + \varepsilon_0 \int_{-\infty}^{\infty} \left[ [\mathbf{E}(\mathbf{r},t) \cdot \mathbf{E}(\mathbf{r},t')] \chi^{(3)}_{nu,b}(t-t') \mathbf{E}(\mathbf{r},t') \right] dt'. \tag{9}$$

Hellwarth, Owyoung and George in [26], called the first term of equation (9) containing $\chi^{(3)}_{nu,a}$ the "isotropic" part of the nonlinear polarization vector due to each vector component of $\mathbf{P}^{NL}_{nu}$ simply being $E_x$, $E_y$ or $E_z$ of $\mathbf{E}(\mathbf{r},t)$ multiplied by the same value, $\varepsilon_0$ times the integral containing $|\mathbf{E}(\mathbf{r},t')|^2$ for a given $(\mathbf{r},t)$, making it isotropic; while the second term of equation (9) containing $\chi^{(3)}_{nu,b}$ was called the "anisotropic" part of the nonlinear polarization vector, due to each vector component of $\mathbf{P}^{NL}_{nu}$, not, simply being $E_x$, $E_y$ or $E_z$ of $\mathbf{E}(\mathbf{r},t)$ multiplied by the same value for a given $(\mathbf{r},t)$. The "anisotropic" aspect can be seen more clearly in equations (20) to (27) and their discussion. This means that using the Born-Oppenheimer approximation in an isotropic media the nonlinear polarization vector can have both isotropic and anisotropic parts, which is not intuitive. This is discussed further in Supplement 1 (Section 1: The polarization vector and rank 4 susceptibility tensor using index notation), along with a tensor description of the equations in this section using index notation for added clarity.

### 2.2 FDTD GVADE method summary: isotropic media with an isotropic polarization vector

The FDTD GVADE method [1-3] does require a few assumptions. First, the Born-Oppenheimer approximation is used with its corresponding assumptions. Second, the linear susceptibility $\chi^{(1)}(\omega)$ is approximated using a Sellmeier pole expansion. This is used in the $\mathbf{J}_{Lorentz}$ calculation, and it is also used to calculate the linear relative permittivity $\varepsilon_r(\omega)$ and the linear refractive index $n_0(\omega) = \sqrt{\varepsilon_r(\omega)}$ which are required for the $\chi^{(3)}(t,\omega_c)$ calculation. Third, the electromagnetic wave is assumed to be operating at a center or carrier frequency, $\omega_c$, where the nonlinear susceptibility can be defined as $\chi^{(3)}(t,\omega_c) = \chi^{(3)}_0(\omega_c)g(t)$, allowing $\chi^{(3)}_0 = (4/3)c\varepsilon_0[n_0(\omega_c)]^2 n_2$, a constant calculated at $\omega_c$, to be pulled out of the nonlinear polarization vector integral [8, 9], where $n_2$ is the third-order nonlinear refractive index and the normalized temporal response function $g(t) = f_{el}\delta_{el}(t) + f_{nu,a}g_{nu,a}(t)$. Fourth, the normalized isotropic nuclear temporal response function, $g_{nu,a}(t)$, has a "closed form analytical" Fourier Transform, or can be approximated well by $g_{nu,a}(t) \approx \sum_m [f_m g_m(t)]$, where $g_m(t)$ have "closed form analytical" Fourier Transforms as described below.

The FDTD GVADE method [1-3] models electromagnetic wave propagation in nonlinear isotropic media at optical frequencies by accounting for the nonlinear behavior through the polarization current terms in Ampere's law. The Ampere's law equation is written at time-step number $n + 1/2$ approximating the time derivative of $\mathbf{E}^{n+1/2}$ using a temporal finite-difference centered at time-step number $n + 1/2$,

$$\nabla \times \mathbf{H}^{n+1/2} = \frac{\varepsilon_0}{\Delta t}(\mathbf{E}^{n+1} - \mathbf{E}^n) + \mathbf{J}_{Lorentz}^{n+1/2} + \mathbf{J}_{el}^{n+1/2} + \mathbf{J}_{nu,a}^{n+1/2}, \tag{10}$$

and is then solved for the electric fields at $n + 1$ time-steps using the multi-dimensional Newton-Raphson method, where the spatial derivatives in the $\nabla \times \mathbf{H}^{n+1/2}$ term are approximated using spatial finite-differences of $\mathbf{H}^{n+1/2}$ centered about the spatial location $(x,y,z)$. Then, using the electric fields at $n + 1$ time-steps as the new electric fields at $n$ time-steps, Faraday's law is solved for the magnetic fields using classic finite-difference time-domain update equations at the new time-step number $n + 1/2$. This is repeated for each new time-step number $n$.

The original FDTD GVADE method [1-3] solves the first convolution integral from equation (9), where $\mathbf{J} = \frac{\partial \mathbf{P}}{\partial t}$, as "auxiliary differential equations" (ADEs) which are solved as finite-difference update equations in Ampere's law using the Newton-Raphson method. As an example of the FDTD GVADE method process for solving a convolution integral using an auxiliary differential equation, the isotropic polarization vector term from equation (9) will be solved following the derivation from [2, 3], starting by defining the convolution integral as a scalar auxiliary variable,

$$S_{nu,a}(t) = \int_{-\infty}^{\infty} \left[\chi_{nu,a}^{(3)}(t - t')\, |\mathbf{E}(\mathbf{r},t')|^2 \right] dt' = \chi_{nu,a}^{(3)}(t) * |\mathbf{E}(t)|^2, \tag{11}$$

which will then be used to create an auxiliary differential equation. Choosing the $g_{nu,a}(t)$ from [3], which models the nuclear isotropic nonlinear behavior in silica using [2, 3, 35-37]

$$\chi_{nu,a}^{(3)}(t) = \chi_0^{(3)} f_{nu,a}\, g_{nu,a}(t), \tag{12}$$

$$g_{nu,a}(t) = \left(\frac{\tau_1^2 + \tau_2^2}{\tau_1 \tau_2^2}\right) e^{-t/\tau_2} \sin(t/\tau_1)\, U(t), \tag{13}$$

with the scalar auxiliary variable,

$$S_{nu,a}(t) = \chi_0^{(3)} f_{nu,a} \left(g_{nu,a}(t) * |\mathbf{E}(t)|^2\right). \tag{14}$$

The resulting auxiliary differential equation is found by taking the Fourier Transform of equation (14), re-arranging and then taking the Inverse Fourier Transform,

$$\omega_{nu,a} \tilde{S}_{nu,a}(t) + 2\delta_{nu,a} \frac{\partial \tilde{S}_{nu,a}(t)}{\partial t} + \frac{\partial^2 \tilde{S}_{nu,a}(t)}{\partial t^2} = f_{nu,a} \chi_0^{(3)} \omega_{nu,a} |\mathbf{E}(t)|^2, \tag{15}$$

where $\omega_{nu,a} = \sqrt{(\tau_1^2 + \tau_2^2)/(\tau_1^2 \tau_2^2)}$ and $\delta_{nu,a} = 1/\tau_2$ [3]. This auxiliary differential equation is solved using a finite central difference equation centered around time-step number $n$, where

$\frac{\partial S}{\partial t} \approx (S^{n+1} - S^{n-1})/(2\Delta t)$ and $\frac{\partial^2 S}{\partial t^2} \approx (S^{n+1} - 2S^n + S^{n-1})/(\Delta t)^2$. Then, the equation is re-arranged to solve for the scalar auxiliary variable at the new time-step number $n + 1$,

$$S_{nu,a}^{n+1} = \left[\frac{2 - \omega_{nu,a}(\Delta t)^2}{\delta_{nu,a}\Delta t + 1}\right] S_{nu,a}^n + \left[\frac{\delta_{nu,a}\Delta t - 1}{\delta_{nu,a}\Delta t + 1}\right] S_{nu,a}^{n-1} + \left[\frac{f_{nu,a}\chi_0^{(3)}\omega_{nu,a}^2(\Delta t)^2}{\delta_{nu,a}\Delta t + 1}\right]|\mathbf{E}^n(t)|^2. \tag{16}$$

The polarization current is then found by numerically differentiating $\mathbf{J} = \frac{\partial \mathbf{P}}{\partial t}$ using a central difference centered around time-step number $n + 1/2$,

$$\mathbf{J}_{nu,a}^{n+1/2} = \frac{\varepsilon_0}{\Delta t}\left(|\mathbf{E}^{n+1}(t)|^2 S_{nu,a}^{n+1}(t) - |\mathbf{E}^n(t)|^2 S_{nu,a}^n(t)\right). \tag{17}$$

The FDTD GVADE method [1-3] solves for the linear polarization current and the nonlinear electronic polarization current as

$$\mathbf{J}_{Lorentz}^{n+1/2} = \frac{1}{2}\sum_{p=1}^{3}\left[(1+\alpha_p)\mathbf{J}_{Lorentz_p}^n - \mathbf{J}_{Lorentz_p}^{n-1} + \frac{\gamma_p}{2\Delta t}\left(\mathbf{E}^{n+1} - \mathbf{E}^{n-1}\right)\right], \tag{18}$$

$$\mathbf{J}_{el}^{n+1/2} = \chi_0^{(3)}f_{el}\frac{\varepsilon_0}{\Delta t}\left\{\left(|\mathbf{E}^{n+1}|\right)^2\mathbf{E}^{n+1} - \left(|\mathbf{E}^n|\right)^2\mathbf{E}^n\right\}. \tag{19}$$

Please see [3] for the full derivation. The FDTD GVADE method [1-3] uses only the isotropic nuclear polarization vector integral in its formulation represented by $\mathbf{J}_{nu,a}^{n+1/2}$. The purpose of this paper is to extend the method to include the anisotropic part of the polarization vector integral into the method using $\mathbf{J}_{nu,b}^{n+1/2}$, where $\mathbf{J}_{nu,b}(t) = \frac{\partial}{\partial t}\mathbf{P}_{nu,b}$.

*2.3 Extended FDTD GVADE method: including the anisotropic part of the polarization vector*

The result of including the anisotropic part of the polarization vector is more required convolution integrals and more auxiliary differential equations. Writing out the vector components of the anisotropic part of the nuclear nonlinear polarization vector, from the anisotropic integral from equation (9), in terms of scalar auxiliary variables, $S_{nu,b,kl}(t)$, where:

$$P_{nu,b,x}^{(3)}(t) = \varepsilon_0\left[E_x(t)S_{nu,b,xx}(t) + E_y(t)S_{nu,b,xy}(t) + E_z(t)S_{nu,b,xz}(t)\right], \tag{20}$$

$$P_{nu,b,y}^{(3)}(t) = \varepsilon_0\left[E_x(t)S_{nu,b,xy}(t) + E_y(t)S_{nu,b,yy}(t) + E_z(t)S_{nu,b,yz}(t)\right], \tag{21}$$

$$P_{nu,b,z}^{(3)}(t) = \varepsilon_0\left[E_x(t)S_{nu,b,xz}(t) + E_y(t)S_{nu,b,yz}(t) + E_z(t)S_{nu,b,zz}(t)\right], \tag{22}$$

where $k = x, y$ or $z$ and $l = x, y$ or $z$ with,

$$S_{nu,b,kl}(t) = \int_{-\infty}^{\infty}[\chi_0^{(3)}f_{nu,b}g_{nu,b}(t - t')\ E_k(t')E_l(t')]dt' = \chi_0^{(3)}f_{nu,b}\left(g_{nu,b}(t)*[E_k(t)E_l(t)]\right). \tag{23}$$

For comparison, the isotropic nuclear polarization vector components are

$$P_{nu,a,x}^{(3)}(t) = \varepsilon_0 E_x(t)\left[S_{nu,a,xx}(t) + S_{nu,a,yy}(t) + S_{nu,a,zz}(t)\right] = E_x(t)S_{nu,a}(t), \tag{24}$$

$$P_{nu,a,y}^{(3)}(t) = \varepsilon_0 E_y(t)\left[S_{nu,a,xx}(t) + S_{nu,a,yy}(t) + S_{nu,a,zz}(t)\right] = E_y(t)S_{nu,a}(t), \tag{25}$$

$$P_{nu,a,z}^{(3)}(t) = \varepsilon_0 E_z(t)\left[S_{nu,a,xx}(t) + S_{nu,a,yy}(t) + S_{nu,a,zz}(t)\right] = E_z(t)S_{nu,a}(t). \tag{26}$$

It can be seen that the "isotropic" part of the polarization vector is defined as isotropic since the electric fields in each Cartesian direction, $E_x, E_y$ and $E_z$ are all multiplied by the same term,

$S_{nu,a}(t)$, as described by $P^{(3)}_{nu,a,i}(t) = \varepsilon_0 E_i(t) S_{nu,a}(t)$ or equivalently $\mathbf{P}^{(3)}_{nu,a}(t) = \varepsilon_0 \mathbf{E}(t) S_{nu,a}(t)$ where $i = x, y$ or $z$. The "anisotropic" term is called anisotropic because $P^{(3)}_{nu,b,i}(t) \neq E_i(t) S_{nu,b}(t)$, but rather a rank 2 tensor $\mathbf{S}_{nu,b}$ multiplied by $\varepsilon_0$, relates $\mathbf{P}^{(3)}_{nu,b}(t)$ and $\mathbf{E}(t)$,

$$\mathbf{P}^{(3)}_{nu,b}(t) = \varepsilon_0 \mathbf{S}_{nu,b} \mathbf{E}(t) = \varepsilon_0 \begin{bmatrix} S_{nu,b,xx}(t) & S_{nu,b,xy}(t) & S_{nu,b,xz}(t) \\ S_{nu,b,xy}(t) & S_{nu,b,yy}(t) & S_{nu,b,yz}(t) \\ S_{nu,b,xz}(t) & S_{nu,b,yz}(t) & S_{nu,b,zz}(t) \end{bmatrix} \begin{bmatrix} E_x(t) \\ E_y(t) \\ E_z(t) \end{bmatrix}. \quad (27)$$

The anisotropic part of the polarization vector can be seen to require six extra convolution integrals, $S_{nu,b,kl}(t)$, for the full 3D case, in addition to the one convolution integral, $S_{nu,a}(t)$, required for the isotropic part of the polarization vector.

The anisotropic polarization current density $\mathbf{J}_{nu,b}(t) = \frac{\partial}{\partial t}\mathbf{P}_{nu,b}$ is found by numerically differentiating equations (20) to (22) using central differencing, centered around time-step number $n + 1/2$ to get the $k$ component of the polarization current,

$$J^{n+1/2}_{nu,b,k} = \frac{\varepsilon_0}{\Delta t}[E_x^{n+1}(t) S^{n+1}_{nu,b,kx}(t) - E_x^n(t) S^n_{nu,b,kx}(t)] + \quad (28)$$
$$\frac{\varepsilon_0}{\Delta t}[E_y^{n+1}(t) S^{n+1}_{nu,b,ky}(t) - E_y^n(t) S^n_{nu,b,ky}(t)] +$$
$$\frac{\varepsilon_0}{\Delta t}[E_z^{n+1}(t) S^{n+1}_{nu,b,kz}(t) - E_z^n(t) S^n_{nu,b,kz}(t)].$$

*2.4 Anisotropic polarization vector convolution integrals and auxiliary differential equations for silica anisotropic temporal response function*

For the anisotropic part of the polarization vector, a similar process to Section 2.2 is used to create and solve the auxiliary differential equations corresponding to equations (20)-(23). As an example, for silica [34, 38],

$$\chi^{(3)}_{nu,b}(t) = \chi^{(3)}_0 [f_b g_b(t) + f_c g_{nu,a}(t)], \quad (29)$$

$$g_b(t) = \left(\frac{2\tau_b - t}{\tau_b^2}\right) e^{-t/\tau_b} U(t) = C_b(2\tau_b - t) e^{-t/\tau_b} U(t), \quad (30)$$

and $C_b = 1/\tau_b^2$. The scalar auxiliary variables are defined as $S_{b,kl}(t) = \chi^{(3)}_0 f_b g_b(t) * [E_k(t) E_l(t)]$ and $S_{c,kl}(t) = \chi^{(3)}_0 f_c g_{nu,a}(t) * [E_k(t) E_l(t)]$, where $S_{nu,b,kl}(t) = S_{b,kl}(t) + S_{c,kl}(t)$. The process is shown below for $S_{b,kl}(t)$, with $S_{b,kl}(t)$ split into two parts $S_{b,kl}(t) = S_{b1,kl}(t) + S_{b2,kl}(t)$,

$$S_{b1,kl}(t) = f_b \chi^{(3)}_0 \left(g_{b,1}(t) * [E_k(t) E_l(t)]\right), \quad (31)$$

$$S_{b2,kl}(t) = f_b \chi^{(3)}_0 \left(g_{b,2}(t) * [E_k(t) E_l(t)]\right), \quad (32)$$

so that closed form analytical Fourier Transforms exist, where $g_{b,1}(t) = C_b(2\tau_b) e^{-t/\tau_b} U(t)$ and $g_{b,2}(t) = -C_b t e^{-t/\tau_b} U(t)$. Taking the Fourier Transform of equation (31), with $\omega_b = 1/\tau_b$,

$$\tilde{S}_{b1,kl}(\omega) = f_b \chi^{(3)}_0 \left(\tilde{g}_{b,1}(\omega) \cdot \mathcal{F}[E_k(t) E_l(t)]\right), \quad (33)$$

with,

$$\tilde{g}_{b,1}(\omega) = C_b \left(\frac{2\tau_b}{\omega_b + j\omega}\right). \quad (34)$$

Equation (34) is plugged into (33), multiplied on both sides by $(\omega_b + j\omega)$ and re-arranged,

$$(\omega_b + j\omega)\tilde{S}_{b1,kl}(\omega) = f_b \chi^{(3)}_0 C_b 2\tau_b \cdot \mathcal{F}[E_k(t) E_l(t)] \cdot \quad (35)$$

The Inverse Fourier Transform is taken, resulting in the $S_{b1,kl}$ auxiliary differential equation,

$$\omega_b S_{b1,kl}(t) + \frac{\partial S_{b1,kl}(t)}{\partial t} = f_b \chi_0^{(3)} C_b 2\tau_b [E_k(t)E_l(t)] . \tag{36}$$

Next, taking the Fourier Transform of equation (32), with again $\omega_b = 1/\tau_b$,

$$\tilde{S}_{b2,kl}(\omega) = f_b \chi_0^{(3)} \left( \tilde{g}_{b,2}(\omega) \cdot \mathcal{F}[E_k(t)E_l(t)] \right), \tag{37}$$

with,

$$\tilde{g}_{b,2}(\omega) = -C_b \left( \frac{1}{(\omega_b + j\omega)^2} \right). \tag{38}$$

Equation (38) is plugged into (37), multiplied on both sides by $(\omega_b + j\omega)^2$ and re-arranged,

$$\left[ \omega_b^2 + 2\omega_b(j\omega) + (j\omega)^2 \right] \tilde{S}_{b2,kl}(w) = -f_b \chi_0^{(3)} C_b \cdot \mathcal{F}[E_k(t)E_l(t)] . \tag{39}$$

The Inverse Fourier Transform is taken, resulting in the $S_{b2,kl}$ auxiliary differential equation,

$$\omega_b^2 S_{b2,kl}(t) + 2\omega_b \frac{\partial S_{b2,kl}(t)}{\partial t} + \frac{\partial^2 S_{b2,kl}(t)}{\partial t^2} = -f_b \chi_0^{(3)} C_b [E_k(t)E_l(t)] . \tag{40}$$

Explicit time-stepping relations are created for equations (36) and (40) using central difference approximations in time about time-step number $n$, where $\frac{\partial S}{\partial t} \approx (S^{n+1} - S^{n-1})/(2\Delta t)$ and $\frac{\partial^2 S}{\partial t^2} \approx (S^{n+1} - 2S^n + S^{n-1})/(\Delta t)^2$, and then each equation is solved for $S^{n+1}$:

$$S_{b1,kl}^{n+1} = [-2\omega_b (\Delta t)] S_{b1,kl}^n + S_{b1,kl}^{n-1} + \left[ 4(\Delta t) f_b \chi_0^{(3)} C_b \tau_b \right] E_k^n(t) E_l^n(t) , \tag{41}$$

$$S_{b2,kl}^{n+1} = \left[ \frac{2 - \omega_b^2 (\Delta t)^2}{\omega_b \Delta t + 1} \right] S_{b2,kl}^n + \left[ \frac{\omega_b \Delta t - 1}{\omega_b \Delta t + 1} \right] S_{b2,kl}^{n-1} + \left[ -\frac{f_b \chi_0^{(3)} C_b (\Delta t)^2}{(\omega_b \Delta t + 1)} \right] E_k^n(t) E_l^n(t) . \tag{42}$$

This process was repeated for $S_{c,kl}(t)$, resulting in the same update equation as equation (16), except $S_{nu,a}^{n+m}$, $|\mathbf{E}^n(t)|^2$ and $f_{nu,a}$ are replaced with $S_{c,kl}^{n+m}$, $[E_k(t)E_l(t)]$ and $f_c$, where $m = 1, 0, -1$.

### 2.5 Polarization vector convolution integrals and auxiliary differential equations for carbon disulfide anisotropic temporal response function

For the isotropic media carbon disulfide ($CS_2$), the nonlinear behavior is composed of the isotropic, "electronic" and nuclear "collision" mechanisms, along with the molecular re-orientational, "librational" and "diffusive" nuclear mechanisms [39, 40]. The "collision" mechanism has isotropic polarization behavior while the molecular re-orientational mechanisms are anisotropic. The $CS_2$ material nonlinear parameters used for this paper follow the model used in [39, 40], while the linear two-pole Sellmeier expansion parameters follow [41], and are listed in Table 3. The isotropic and anisotropic parts of the nuclear nonlinear susceptibility for $CS_2$ are,

$$\chi_{nu,a}^{(3)}(t) = \chi_0^{(3)} f_{nu,a} g_{nu,a}(t) = \chi_0^{(3)} f_c g_c(t) , \tag{43}$$

$$\chi_{nu,b}^{(3)}(t) = \chi_0^{(3)} f_{nu,b} g_{nu,b}(t) = \chi_0^{(3)} [f_d g_d(t) + f_v g_v(t)] , \tag{44}$$

where subscripts $c$, $d$ and $v$ correspond to "collision", "diffusive" and "libration" for $CS_2$. The nuclear collision and nuclear diffusive parts of the nonlinear behavior of $CS_2$ are modeled by the normalized overdamped functions $g_c(t)$ and $g_d(t)$ [39],

$$g_c(t) = C_c \left( 1 - e^{-\frac{t}{\tau_{r,c}}} \right) e^{-\frac{t}{\tau_{f,c}}} U(t) = C_c \left( e^{-\frac{t}{\tau_{f,c}}} - e^{-t\left(\frac{1}{\tau_{r,c}} + \frac{1}{\tau_{f,c}}\right)} \right) U(t) , \tag{45}$$

$$g_d(t) = C_d\left(1 - e^{-\frac{t}{\tau_{r,d}}}\right)e^{-\frac{t}{\tau_{f,d}}} U(t) = C_d\left(e^{-\frac{t}{\tau_{f,d}}} - e^{-t\left(\frac{1}{\tau_{r,d}}+\frac{1}{\tau_{f,d}}\right)}\right)U(t), \quad (46)$$

where $C_c = (\tau_{r,c} + \tau_{f,c})/\tau_{f,c}^2$ and $C_d = (\tau_{r,d} + \tau_{f,d})/\tau_{f,d}^2$. The auxiliary differential equation for the overdamped response function is created and solved for the diffusive function as an example. The Fourier Transform of equation (46) is taken, where $\omega_{f,d} = \frac{1}{\tau_{f,d}}$ and $\omega_{r,f,d} = \frac{1}{\tau_{r,d}} + \frac{1}{\tau_{f,d}}$,

$$\tilde{S}_{d,kl}(\omega) = f_d \chi_0^{(3)} \left(\tilde{g}_d(\omega) \cdot \mathcal{F}[E_k(t)E_l(t)]\right), \quad (47)$$

where,

$$\tilde{g}_d(\omega) = C_d\left(\frac{1}{\omega_{f,d}+j\omega} - \frac{1}{\omega_{r,f,d}+j\omega}\right). \quad (48)$$

Equation (48) is plugged into equation (47), multiplied on both sides by $(\omega_{f,d}+j\omega)(\omega_{r,f,d}+j\omega)$ and re-arranged,

$$[\omega_{f,d}\omega_{r,f,d} + (\omega_{f,d}+\omega_{r,f,d})(j\omega) + (j\omega)^2]\tilde{S}_{d,kl}(\omega) = \quad (49)$$
$$f_d\chi_0^{(3)}C_d[\omega_{r,f,d} - \omega_{f,d}] \cdot \mathcal{F}[E_k(t)E_l(t)].$$

The inverse Fourier transform is taken, resulting in the auxiliary differential equation,

$$\omega_{f,d}\omega_{r,f,d} S_{d,kl}(t) + [\omega_{f,d}+\omega_{r,f,d}]\frac{\partial S_{d,kl}(t)}{\partial t} + \frac{\partial^2 S_{d,kl}(t)}{\partial t^2} = \quad (50)$$
$$f_d\chi_0^{(3)}C_d[\omega_{r,f,d} - \omega_{f,d}] [E_k(t)E_l(t)],$$

Explicit time-stepping relations are created for equation (50) using central difference approximations in time about time-step number $n$, where $\frac{\partial S}{\partial t} \approx (S^{n+1} - S^{n-1})/(2\Delta t)$ and $\frac{\partial^2 S}{\partial t^2} \approx (S^{n+1} - 2S^n + S^{n-1})/(\Delta t)^2$, and then solved for $S_{d,kl}^{n+1}$:

$$S_{d,kl}^{n+1} = \frac{2(2 - (\Delta t)^2 \omega_{f,d}\omega_{r,f,d})}{(2+\Delta t[\omega_{f,d}+\omega_{r,f,d}])}S_{d,kl}^n + \frac{(\Delta t[\omega_{f,d}+\omega_{r,f,d}]-2)}{(2+\Delta t[\omega_{f,d}+\omega_{r,f,d}])}S_{d,kl}^{n-1} +$$
$$\frac{2(\Delta t)^2 f_d\chi_0^{(3)}C_d[\omega_{r,f,d}-\omega_{f,d}]}{(2+\Delta t[\omega_{f,d}+\omega_{r,f,d}])}[E_k(t)E_l(t)]. \quad (51)$$

This auxiliary variable update equation for $S_{d,kl}^{n+1}$, is then used along with the "librational" auxiliary variable $S_{v,kl}^{n+1}$ update equation to determine $S_{nu,b,kl}^{n+1}$ using $S_{nu,b,kl}^{n+1} = S_{d,kl}^{n+1} + S_{v,kl}^{n+1}$, which is then used in equation (28) to determine $\mathbf{J}_{nu,b}^{n+1/2}$.

The "librational" temporal response function of $CS_2$, is modelled by an underdamped quantum harmonic oscillator [35, 39],

$$g_{v,\text{Quantum}}(t) = C_v e^{-\frac{t}{\tau_{f,v}}} U(t) \int_0^\infty \frac{\sin(\omega t)}{\omega} h(\omega)d\omega. \quad (52)$$

$$h(\omega) = e^{-\frac{(\omega-\omega_v)^2}{2\sigma^2}} - e^{-\frac{(\omega+\omega_v)^2}{2\sigma^2}}. \quad (53)$$

However, there is not a simple closed form Fourier Transform of this model, which is required by the FDTD GVADE method, and as a result, an approximation of that function was required to be found. While more complicated and potentially more accurate approximations could have been used, for example the sum of underdamped, critically and overdamped oscillator solutions [35, 37], the classical underdamped oscillator model, was chosen as a simple proof of concept:

$$g_v(t) = C_v e^{-t/\tau_{v,2}} \sin(t/\tau_{v,1}) U(t), \quad (54)$$

$$C_v = \left(\frac{\tau_{v,1}^2 + \tau_{v,2}^2}{\tau_{v,1}\tau_{v,2}^2}\right). \tag{55}$$

The normalized libration temporal response function $g_v(t)$ from [39] and our approximation are plotted versus time scaled by $f_v$, along with the other normalized temporal response functions from [39] scaled by $f_m$ and their summations, in Fig. 1, where $m = c, l, d$. It can be seen visually that the classical underdamped solution is a reasonably good fit for the quantum harmonic oscillator model temporal response function, and the auxiliary differential equation solution is already known.

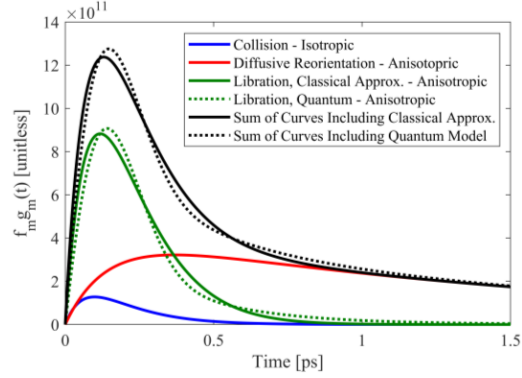

Fig. 1. For comparison purposes, the normalized temporal response functions of $CS_2$ [39], our approximation of the quantum model of the libration temporal response function and their summations are plotted verses time, similar to Fig. 2 in [39], with $\tau_{v,1}$=525.21fs and $\tau_{v,2}$=119.7fs.

## 2.6 Solving Ampere's Law for the electric fields using the Newton-Raphson method

Following and extending the approach from [1-3], the electric field at time-step number $n + 1$ is determined using the Newton-Raphson method to solve Ampere's law at $n + 1/2$ time-steps,

$$\nabla \times \mathbf{H}^{n+1/2} = \frac{\varepsilon_0}{\Delta t}(\mathbf{E}^{n+1} - \mathbf{E}^n) + \mathbf{J}_{\text{Lorentz}}^{n+1/2} + \mathbf{J}_{\text{el}}^{n+1/2} + \mathbf{J}_{\text{nu,a}}^{n+1/2} + \mathbf{J}_{\text{nu,b}}^{n+1/2}. \tag{56}$$

Plugging in the polarization current expressions into Ampere's law at $n + \tfrac{1}{2}$ time-steps, results in an equation which can be used as part of the Newton-Raphson method to determine $\mathbf{E}^{n+1}$,

$$\begin{aligned}
\begin{bmatrix}X\\Y\\Z\end{bmatrix} &= -\nabla \times \mathbf{H}^{n+1/2} + \frac{\varepsilon_0}{\Delta t}(\mathbf{E}^{n+1} - \mathbf{E}^n) + \\
&\quad \frac{1}{2}\sum_{p=1}^{3}\left[(1+\alpha_p)\mathbf{J}_{\text{Lorentz}_p}^n - \mathbf{J}_{\text{Lorentz}_p}^{n-1} + \frac{\gamma_p}{2\Delta t}(\mathbf{E}^{n+1} - \mathbf{E}^{n-1})\right] + \\
&\quad \chi_0^{(3)} f_{\text{el}}\frac{\varepsilon_0}{\Delta t}\left\{(|\mathbf{E}^{n+1}|)^2\mathbf{E}^{n+1} - (|\mathbf{E}^n|)^2\mathbf{E}^n\right\} + \frac{\varepsilon_0}{\Delta t}(\mathbf{E}^{n+1}S_{\text{nu,a}}^{n+1} - \mathbf{E}^n S_{\text{nu,a}}^n) + \\
&\quad \frac{\varepsilon_0}{\Delta t}(\mathbf{S}_{\text{nu,b}}^{n+1}\mathbf{E}^{n+1} - \mathbf{S}_{\text{nu,b}}^n\mathbf{E}^n).
\end{aligned} \tag{57}$$

The Newton-Raphson method solves for the electric field by iterative guessing based on the Jacobian Matrix, until $X$, $Y$ and $Z$ become sufficiently close zero, using [42]

$$\begin{bmatrix}E_x^{n+1}\\E_y^{n+1}\\E_z^{n+1}\end{bmatrix}_{g+1} = \begin{bmatrix}E_x^{n+1}\\E_y^{n+1}\\E_z^{n+1}\end{bmatrix}_g - \left(\mathbf{M}^{-1}\begin{bmatrix}X\\Y\\Z\end{bmatrix}\right)\Bigg|_g. \tag{58}$$

Each Newton-Raphson iteration guess values of $E_x^{n+1}$, $E_y^{n+1}$ and $E_z^{n+1}$ at guess number "g", are used to then estimate the next iteration values of $E_x^{n+1}$, $E_y^{n+1}$ and $E_z^{n+1}$ at guess number "g+1". The Jacobian matrix **M** is defined by $\partial(X,Y,Z)/\partial(E_x^{n+1},E_y^{n+1},E_z^{n+1})$,

$$M_{11} = \frac{\varepsilon_0}{\Delta t}\left[1 + f_{el}\chi_0^{(3)}\left\{3(E_x^{n+1})^2 + (E_y^{n+1})^2 + (E_z^{n+1})^2\right\} + S_{nu,a}^{n+1} + S_{nu,b,xx}^{n+1}\right] \quad (59)$$
$$+ \frac{1}{4\Delta t}(\gamma_1 + \gamma_2 + \gamma_3),$$

$$M_{22} = \frac{\varepsilon_0}{\Delta t}\left[1 + f_{el}\chi_0^{(3)}\left\{(E_x^{n+1})^2 + 3(E_y^{n+1})^2 + (E_z^{n+1})^2\right\} + S_{nu,a}^{n+1} + S_{nu,b,yy}^{n+1}\right] \quad (60)$$
$$+ \frac{1}{4\Delta t}(\gamma_1 + \gamma_2 + \gamma_3),$$

$$M_{33} = \frac{\varepsilon_0}{\Delta t}\left[1 + f_{el}\chi_0^{(3)}\left\{(E_x^{n+1})^2 + (E_y^{n+1})^2 + 3(E_z^{n+1})^2\right\} + S_{nu,a}^{n+1} + S_{nu,b,zz}^{n+1}\right] \quad (61)$$
$$+ \frac{1}{4\Delta t}(\gamma_1 + \gamma_2 + \gamma_3),$$

$$M_{12} = M_{21} = f_{el}\frac{\varepsilon_0\chi_0^{(3)}}{\Delta t}2E_x^{n+1}E_y^{n+1} + \frac{\varepsilon_0}{\Delta t}(S_{nu,b,xy}^{n+1}), \quad (62)$$

$$M_{13} = M_{31} = f_{el}\frac{\varepsilon_0\chi_0^{(3)}}{\Delta t}2E_x^{n+1}E_z^{n+1} + \frac{\varepsilon_0}{\Delta t}(S_{nu,b,xz}^{n+1}), \quad (63)$$

$$M_{23} = M_{32} = f_{el}\frac{\varepsilon_0\chi_0^{(3)}}{\Delta t}2E_y^{n+1}E_z^{n+1} + \frac{\varepsilon_0}{\Delta t}(S_{nu,b,yz}^{n+1}). \quad (64)$$

Note the off-diagonal terms of the Jacobian now have an extra term corresponding to, $S_{nu,b,xy}^{n+1}$, $S_{nu,b,xz}^{n+1}$, $S_{nu,b,yz}^{n+1}$, that was not present in the isotropic polarization vector FDTD GVADE method, along with the diagonal terms adding $S_{nu,b,xx}^{n+1}$, $S_{nu,b,yy}^{n+1}$, $S_{nu,b,zz}^{n+1}$.

## 2.7 Special case: 2D transverse magnetic to z

An important special case, transverse magnetic to z, is simpler than examples requiring the full 3D Maxwell's equations. However, it does still include a few components of the anisotropic part of the polarization vector, allowing for an anisotropic case to be simulated while requiring less computational resources. In the case where the field components do not depend on the $y$ coordinate, this causes the $\frac{\partial}{\partial y}$ terms from Maxwell's equations to be zero, leaving only the $H_y$, $E_x$ and $E_z$ fields,

$$(\nabla \times \mathbf{E})_{y,2D} = \frac{\partial E_x}{\partial z} - \frac{\partial E_z}{\partial x} = -\mu_0 \frac{\partial H_y}{\partial t}, \quad (65)$$

$$(\nabla \times \mathbf{H})_{x,2D} = -\frac{\partial H_y}{\partial z} = \varepsilon_0 \frac{\partial E_x}{\partial t} + J_x, \quad (66)$$

$$(\nabla \times \mathbf{H})_{z,2D} = \frac{\partial H_y}{\partial x} = \varepsilon_0 \frac{\partial E_z}{\partial t} + J_z. \quad (67)$$

The nonlinear polarization vector $\mathbf{P}^{(3)}(t)$ simplifies, but it still includes anisotropic terms,

$$P_x^{NL}(t) = \varepsilon_0\left[\chi_{el}^{(3)}E_x(t)|\mathbf{E}(t)|^2 + E_x(t)S_{nu,a}(t) + E_x(t)S_{nu,b,xx}(t) + E_z(t)S_{nu,b,xz}(t)\right], \quad (68)$$

$$P_z^{NL}(t) = \varepsilon_0\left[\chi_{el}^{(3)}E_z(t)|\mathbf{E}(t)|^2 + E_z(t)S_{nu,a}(t) + E_x(t)S_{nu,b,xz}(t) + E_z(t)S_{nu,b,zz}(t)\right]. \quad (69)$$

The Newton-Raphson method simplifies to $\mathbf{E}^{n+1} = \hat{\mathbf{a}}_x E_x^{n+1} + \hat{\mathbf{a}}_z E_z^{n+1}$,

$$\begin{bmatrix} X \\ Z \end{bmatrix} = -\nabla \times \mathbf{H}^{n+1/2} + \frac{\varepsilon_0}{\Delta t}(\mathbf{E}^{n+1} - \mathbf{E}^n) +$$
$$\frac{1}{2}\sum_{p=1}^{3}\left[(1+\alpha_p)\mathbf{J}^n_{\text{Lorentz}_p} - \mathbf{J}^{n-1}_{\text{Lorentz}_p} + \frac{\gamma_p}{2\Delta t}(\mathbf{E}^{n+1} - \mathbf{E}^{n-1})\right] +$$
$$\chi_0^{(3)} f_{el}\frac{\varepsilon_0}{\Delta t}\{(|\mathbf{E}^{n+1}|)^2\mathbf{E}^{n+1} - (|\mathbf{E}^n|)^2\mathbf{E}^n\} + \frac{\varepsilon_0}{\Delta t}(\mathbf{E}^{n+1}S^{n+1}_{\text{nu},a} - \mathbf{E}^n S^n_{\text{nu},a})$$
$$+ \begin{bmatrix} \frac{\varepsilon_0}{\Delta t}(E_x^{n+1}S^{n+1}_{\text{nu},b,xx} - E_x^n S^n_{\text{nu},b,xx}) + \frac{\varepsilon_0}{\Delta t}(E_z^{n+1}S^{n+1}_{\text{nu},b,xz} - E_z^n S^n_{\text{nu},b,xz}) \\ \frac{\varepsilon_0}{\Delta t}(E_x^{n+1}S^{n+1}_{\text{nu},b,xz} - E_x^n S^n_{\text{nu},b,xz}) + \frac{\varepsilon_0}{\Delta t}(E_z^{n+1}S^{n+1}_{\text{nu},b,zz} - E_z^n S^n_{\text{nu},b,zz}) \end{bmatrix},$$
(70)

where,

$$\begin{bmatrix} E_x^{n+1} \\ E_z^{n+1} \end{bmatrix}_{g+1} = \begin{bmatrix} E_x^{n+1} \\ E_z^{n+1} \end{bmatrix}_g - \left(\mathbf{M}_{\text{TM}}^{-1}\begin{bmatrix} X \\ Z \end{bmatrix}\right)\bigg|_g, \quad (71)$$

$$M_{\text{TM},11} = \frac{1}{4\Delta t}\sum_{p=1}^{3}\gamma_p + \frac{\varepsilon_0}{\Delta t}\left[1 + f_{el}\chi_0^{(3)}\{3(E_x^{n+1})^2 + (E_z^{n+1})^2\} + S^{n+1}_{\text{nu},a} + S^{n+1}_{\text{nu},b,xx}\right], \quad (72)$$

$$M_{\text{TM},22} = \frac{1}{4\Delta t}\sum_{p=1}^{3}\gamma_p + \frac{\varepsilon_0}{\Delta t}\left[1 + f_{el}\chi_0^{(3)}\{(E_x^{n+1})^2 + 3(E_z^{n+1})^2\} + S^{n+1}_{\text{nu},a} + S^{n+1}_{\text{nu},b,zz}\right], \quad (73)$$

$$M_{\text{TM},12} = M_{\text{TM},21} = f_{el}\frac{\varepsilon_0\chi_0^{(3)}}{\Delta t}2E_x^{n+1}E_z^{n+1} + \frac{\varepsilon_0}{\Delta t}(S^{n+1}_{\text{nu},b,xz}). \quad (74)$$

Note the off-diagonal terms of the Jacobian now have an extra term corresponding to, $S^{n+1}_{\text{nu},b,xz}$, that was not present in the isotropic polarization vector FDTD GVADE method.

*2.8 Special case: transverse electric to z*

Another important special case is the transverse electric to z. This case is simpler than the full 3D Maxwell's equations, removing all anisotropic components of the polarization vector due to only having one electric field component, matching the isotropic polarization vector FDTD GVADE method. In the case where the field components do not depend on the $y$ coordinate, the $\frac{\partial}{\partial y}$ terms from Maxwell's equations become zero, leaving only the $E_y$, $H_x$ and $H_z$ fields,

$$(\nabla \times \mathbf{H})_{y,2D} = \frac{\partial H_x}{\partial z} - \frac{\partial H_z}{\partial x} = \varepsilon_0\frac{\partial E_y}{\partial t} + J_y, \quad (75)$$

$$(\nabla \times \mathbf{E})_{x,2D} = \frac{\partial E_y}{\partial z} = \mu_0\frac{\partial H_x}{\partial t}, \quad (76)$$

$$(\nabla \times \mathbf{E})_{z,2D} = \frac{\partial E_y}{\partial x} = -\mu_0\frac{\partial H_z}{\partial t}. \quad (77)$$

The nonlinear polarization vector $\mathbf{P}^{(3)}(t)$ simplifies greatly, removing all anisotropic terms,

$$P_y^{\text{NL}}(t) = \varepsilon_0[\chi_{el}^{(3)}E_y(t)[E_y(t)]^2 + E_y(t)S_{\text{nu},a}(t) + E_y(t)S_{\text{nu},b,yy}(t)], \quad (78)$$

$$P_y^{\text{NL}}(t) = \quad (79)$$
$$\varepsilon_0\chi_0^{(3)}E_y(t)\int_{-\infty}^{\infty}\left[f_{el}\delta(t-t') + f_{\text{nu},a}g_{\text{nu},a}(t-t') + f_{\text{nu},b}g_{\text{nu},b}(t-t')\right]E_y^2(t')\,dt',$$

which matches the simple example in [34], except this example is *y*-polarized instead of *x*-polarized.

The Newton-Raphson method simplifies to $\mathbf{E}^{n+1} = \hat{\mathbf{a}}_y E_y^{n+1}$,

$$[Y] = -\nabla\times\mathbf{H}^{n+1/2} + \frac{\varepsilon_0}{\Delta t}(\mathbf{E}^{n+1} - \mathbf{E}^n) + \frac{1}{2}\sum_{p=1}^{3}\left[(1+\alpha_p)\mathbf{J}_{\text{Lorentz}_p}^n - \mathbf{J}_{\text{Lorentz}_p}^{n-1} + \frac{\gamma_p}{2\Delta t}(\mathbf{E}^{n+1} - \mathbf{E}^{n-1})\right] + \chi_0^{(3)} f_{\text{el}} \frac{\varepsilon_0}{\Delta t}\{(|\mathbf{E}^{n+1}|)^2 \mathbf{E}^{n+1} - (|\mathbf{E}^n|)^2 \mathbf{E}^n\} + \frac{\varepsilon_0}{\Delta t}(\mathbf{E}^{n+1} S_{\text{nu,a}}^{n+1} - \mathbf{E}^n S_{\text{nu,a}}^n) + \frac{\varepsilon_0}{\Delta t}(\mathbf{E}^{n+1} S_{\text{nu,b},yy}^{n+1} - \mathbf{E}^n S_{\text{nu,b},yy}^n),$$

(80)

where,

$$[E_y^{n+1}]_{g+1} = [E_y^{n+1}]_g - (M_{\text{TE}}^{-1}[Y])|_g,$$

(81)

$$M_{\text{TE}} = \frac{1}{4\Delta t}\sum_{p=1}^{3}\gamma_p + \frac{\varepsilon_0}{\Delta t}[1 + f_{\text{el}}\chi_0^{(3)}\{3(E_y^{n+1})^2\} + S_{\text{nu,a}}^{n+1} + S_{\text{nu,b},yy}^{n+1}].$$

(82)

Note how much simpler this version of the Newton-Raphson is than the prior two sections, since, due to only one electric field component there are no off diagonal terms.

### 3. Numerical results and analysis

In this section, 2D $TM_z$ electromagnetic waves propagating in nonlinear isotropic materials are numerically simulated. For the first example, the original FDTD GVADE 2D $TM_z$ example from [1, 2] is revisited using the nonlinear response tensor from [34, 38] for fused silica and the linear response and simulation parameters from [1, 2]. For the second example, a 2D $TM_z$ simulation was performed in $CS_2$ including its molecular re-orientation mechanisms as the anisotropic part of the nonlinear polarization vector. The goal of these simulations was to show simple examples and proof-of-concepts, simulating an electromagnetic wave propagating in a nonlinear isotropic material with both isotropic and anisotropic parts of the polarization vector, illustrating how the extended FDTD GVADE method is implemented.

All the computational simulation results in this article were conducted with the advanced computing resources provided by Texas A&M High Performance Research Computing.

The 2D $TM_z$ fields were simulated using a *nx* x *nz* dimension Lebedev grid [4, 14, 43], in the same manner as [44], except in this paper the simulation space was not cut in half using symmetry. Liu's paper [43] calls this grid the "unstaggered grid", which is a combination of two shifted Yee grids resulting in collocated electric field components. The two Yee grids are coupled through the nonlinear polarization current density terms.

*3.1 Simulation results: material and parameters - silica*
The material parameters for fused silica from [38] are listed in Table 1, where $\varepsilon_r(\omega_c) = [n_0(\omega_c)]^2 = 2.15188$ and $\chi_0^{(3)} = (4/3)c\,\varepsilon_0\varepsilon_r n_2 = 1.98\times10^{-22}\text{m}^2/\text{V}^2$. For clarity, the variables $f_a, f_b$ and $f_c$ are written as products along with their actual value, to avoid any confusion since [38] defines $f_a, f_b$ and $f_c$ differently than this paper, but the response tensor is equivalent, where $f_m = f_R f_{m,[38]}$ with $m = a,b,c$ and $g_{\text{nu,a}}(t) = g_a(t)$ and $f_{\text{nu,a}} = f_a$. See

Supplement 1 (Section 2: Silica rank 4 susceptibility tensor) for the rank 4 material response tensor written using index notation. The simulation parameters are listed in Table 2, and the time-step, $\Delta t$, was chosen empirically by performing numerical studies as 7.527 times below the required time-step for the Courant stability limit. The vacuum carrier wavelength was selected as $\lambda_0$=433 nm to match [44], and the corresponding material linear wavelength of $\lambda_d \approx 295.2$ nm (resulting in a 55:1 resolution following [3] using 5.333333nm square grid cells). For the purely isotropic polarization vector case, the anisotropic part of the polarization vector is removed, where $f_{R,\text{Isotropic}}$ is slightly different then $f_R$ as described by [34, 38],

$$\mathbf{P}^{\text{NL}}_{\text{Isotropic}} = \varepsilon_0 \chi_0^{(3)} \mathbf{E}(\mathbf{r},t) \int_{-\infty}^{\infty} \left[ g^{(3)}_{\text{Isotropic}}(t-t') |\mathbf{E}(\mathbf{r},t')|^2 \right] dt', \tag{83}$$

$$g^{(3)}_{\text{Isotropic}}(t) = (1 - f_{R,\text{Isotropic}})\delta(t) + f_{R,\text{Isotropic}} g_{\text{nu},a}(t). \tag{84}$$

**Table 1. Material Parameters**

| Parameter | Value |
|---|---|
| $\beta_1$ | 0.69617 |
| $\beta_2$ | 0.40794 |
| $\beta_3$ | 0.89748 |
| $\omega_1$ | $2.7537 \times 10^{16}$ rad/s |
| $\omega_2$ | $1.6205 \times 10^{16}$ rad/s |
| $\omega_3$ | $1.9034 \times 10^{14}$ rad/s |
| $n_2$ | $2.6 \times 10^{-20}$ m$^2$/W |
| $\tau_1$ | 12.2 fs |
| $\tau_2$ | 32 fs |
| $\tau_b$ | 96 fs |
| $f_{el}$ | (1 - 0.245) = 0.755 |
| $f_a$ | (0.245)(0.75) = 0.18375 |
| $f_b$ | (0.245)(0.21) = 0.05145 |
| $f_c$ | (0.245)(0.04) = 0.0098 |
| $f_R$ | 0.245 |
| $f_{R,\text{Isotropic}}$ | 0.18 |

*From [1, 2, 34, 38, 45]*

**Table 2. Simulation Parameters**

| Parameter | Value |
|---|---|
| $\Delta z = \Delta x$ | $5.333333 \times 10^{-9}$ meters |
| $\Delta t$ | $3.34 \times 10^{-18}$ seconds |
| $nz$ | 22501 |
| $nx$ | 5001 |
| $n$ | 72,000 time-steps |
| $\omega_c$ | $4.35 \times 10^{15}$ rad/s ($\lambda_0$~433nm) |
| $w_0$ | 667 nm |
| $A_0$ | $4.77 \times 10^7$ A/m |

*Some Values Selected From [1, 2, 44]*

### 3.2 Simulation results: material and parameters - CS$_2$

The material nonlinear parameters used for CS$_2$ follow the model used in [39, 40], while the linear two-pole Sellmeier expansion parameters follow [41], and are listed in Table 3, where $\varepsilon_r(\omega_c) = [n_0(\omega_c)]^2 = 2.60488$. The electronic and nuclear parts of the susceptibility can be calculated using $\chi^{(3)}_{el} = (4/3) c \varepsilon_0 \varepsilon_r n_{2,el}$ and $\chi^{(3)}_{nu}(t) = 2 c \varepsilon_0 \varepsilon_r R(t)$, where $R(t) = \sum_m [n_{2,m} g_m(t)]$ is as described in [39, 46], with $\chi_0^{(3)} = 3.81677 \times 10^{-20}$ m$^2$/V$^2$, $f_{el} = 0.0362$, $f_c = 0.0362$, $f_l = 0.2754$ and $f_d = 0.65217$, where $g_{\text{nu},a}(t) = g_c(t)$ and $f_{\text{nu},a} = f_c$. Note that $g_c(t)$ and $f_c$ are different than for silica. See Supplement 1 (Section 2: Carbon disulfide rank 4 susceptibility tensor) for the rank 4 material response tensor written using index notation, and for details on how $\chi_0^{(3)}, f_{el}, f_c$, $f_l$ and $f_d$ were determined from the Table 3 parameters. The simulation parameters are listed in Table 4, and the time-step, $\Delta t$, was chosen as 7.051 times below the required time-step for the Courant stability limit. The vacuum carrier wavelength was selected at $\lambda_0 = 694.3$ nm for a ruby laser, and the corresponding material linear wavelength of $\lambda_d \approx 430$ nm (resulting in a 54:1 resolution following [3] using 8nm square grid cells).

Table 3. CS$_2$ Material Parameters

| Parameter | Value |
|---|---|
| $\beta_1$ | 1.499426 |
| $\beta_2$ | 0.089531 |
| $\omega_1$ | $2\pi(c/\lambda_1) =$ 1.0537145×10$^{16}$ rad/s |
| $\omega_2$ | $2\pi(c/\lambda_2) =$ 2.8575045×10$^{14}$ rad/s |
| $\tau_{r,c}$ | 150 ± 50 fs |
| $\tau_{f,c}$ | 140 ± 50 fs |
| $\tau_{r,d}$ | 150 ± 50 fs |
| $\tau_{f,d}$ | 1610 ± 50 fs |
| $\tau_{v,1-approx}$ | 525.21 fs |
| $\tau_{v,2-approx}$ | 119.7 fs |
| $n_{2,el}$ | $(1.5 \pm 0.4) \times 10^{-19}$ m$^2$/W |
| $n_{2,c}$ | $(1.0 \pm 0.2) \times 10^{-19}$ m$^2$/W |
| $n_{2,v}$ | $(7.6 \pm 1.5) \times 10^{-19}$ m$^2$/W |
| $n_{2,d}$ | $(18 \pm 3) \times 10^{-19}$ m$^2$/W |

From References[39-41]

Table 4. CS$_2$ Simulation Parameters

| Parameter | Value |
|---|---|
| $\Delta z = \Delta x$ | $8 \times 10^{-9}$ meters |
| $\Delta t$ | 5.3486×10$^{-18}$ seconds |
| $nz$ | 25001 |
| $nx$ | 3001 |
| $n$ | 55,000 time-steps |
| $\omega_c$ | 2.713 ×10$^{15}$ rad/s ($\lambda_0 \sim 694.3$nm) |
| $w_0$ | 2,000 nm |
| $A_0$ | 4.29×10$^6$ A/m |

Some Values Selected From [44]

### 3.3 Simulation results and discussion

The hard source used to excite the electromagnetic wave at $z = 0$ is [2],

$$H_y(t) = A_0 \sin(\omega_c t) \operatorname{sech}(x/w_0) . \tag{85}$$

Simulation results of $|\mathbf{E}|$ for a 2D TM$_z$ electromagnetic wave in fused silica, using both the original FDTD GVADE isotropic polarization vector model and the extended FDTD GVADE method presented in this paper are shown in Fig. 2, along with $H_y(t, x = 0)$ versus propagation distance $z$ curves for both cases, showing the similarity between the simulation results. The normalized half-peak-input-power beamwidth (HPIPBW) and normalized magnitude of the magnetic field, $|H_y(z)|$, at extremum locations curves are plotted versus propagation distance in Fig. 3 at time $t = (\Delta t)n = 240.5$ fs. Also, simulation results for input amplitudes of $1.02A_0$ and $1.05A_0$ are shown in Fig. 3 as well, with the polarization vector including an anisotropic part, for comparison purposes. A "moving-mean" was used to smooth the curves allowing the behavior to be more easily seen and compared, while maintaining the shape of the curves.

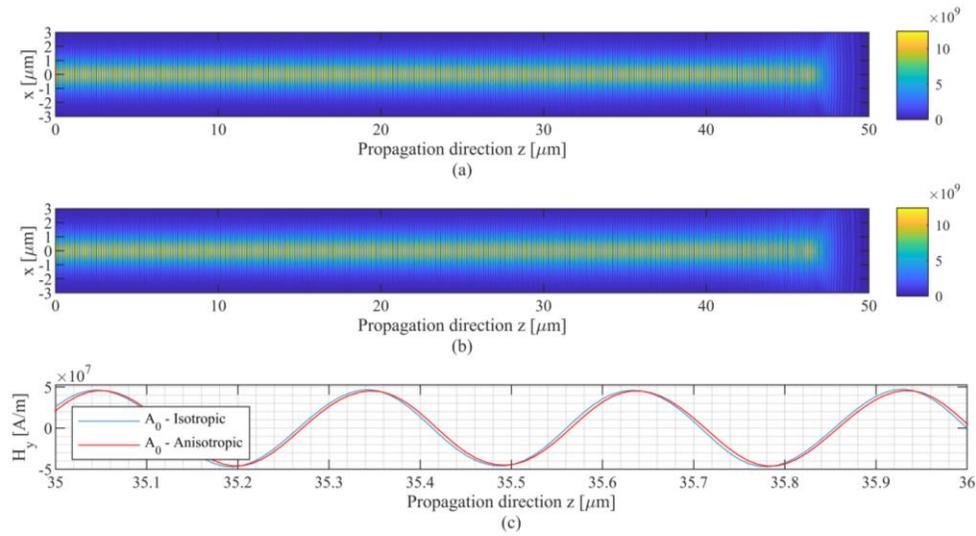

Fig. 2. Results for the 2D TM$_z$ simulation with input amplitude $A_0$ in silica at time $t = (\Delta t)n = 240.5$ fs. (a) |**E**| for the original FDTD GVADE - Isotropic Polarization Vector. To avoid confusion, please see Figure 6.2 from [1] for comparison, rather than [2]. Note that [1, 2] use slightly different nonlinear parameters than this paper. (b) |**E**| for the extended FDTD GVADE – Including Anisotropic Polarization Vector. (c) $H_y(x = 0)$ for (a) and (b) plotted as blue and red curves respectively.

Note that in Fig. 3, the HPIPBW and $|H_y(z)|$ at the extremums between the source at $z = 0$ and the front section of the electromagnetic wave are fairly constant with respect to $z$, with only small variations, but the HPIPBW and $|H_y(z)|$ near the front of the wave show more variation with respect to $z$. This "transient" behavior at the front of the wave is due to the convolution in time of the time-dependent material susceptibility with the electric fields at a given point in space as described in equations (4) and (9); this results in non-instantaneous changes of the polarization vector at that point in space, and the polarization vector approaches "steady state" behavior at that point in space after the front section of the wave propagates beyond it. By observation of the simulation results shown in Fig. 3, the extended version with the polarization vector including the anisotropic part appears to require a slightly larger input amplitude to produce the same HPIPBW as the wave propagates. Otherwise, the simulation results seem to be very similar for fused silica. While this is a simple example, it shows that the isotropic polarization vector model and the more accurate model including the anisotropic part of the polarization vector results vary slightly.

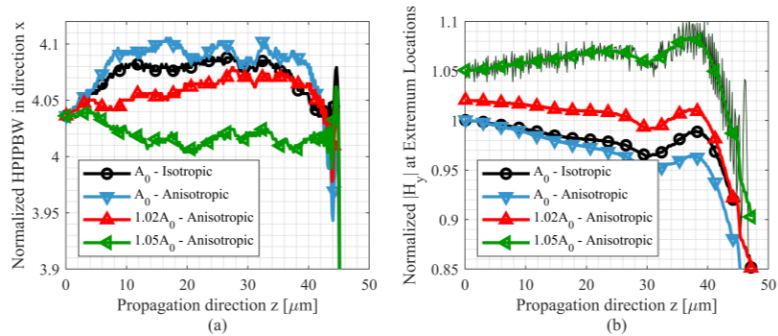

Fig. 3. Smoothed normalized HPIPBW($z$) and normalized $|H_y(z)|$ versus propagation distance $z$, at the extremum locations in the propagation direction $z$ resulting from the source term $\sin(\omega_c t)$, in silica, at time $t = (\Delta t)n = 240.5$ fs. (a) Smoothed HPIPBW($z$) calculated relative to

$\max(|H_y(z = z_{nearest})|)$, normalized to the approximate material wavelength $\lambda_d = \lambda_0/\sqrt{\varepsilon_r}$, (b) Smoothed $|H_y(z)|$ at the extremum location, normalized to the $|H_y(z = z_{nearest})|$, where $z_{nearest}$ is the nearest extremum location to $z = 0$. Note: The thinner darker green curve is the "unsmoothed" $1.05A_0$ input amplitude case, illustrating the "moving-mean" smoothing process used for each curve in the figure.

In contrast, the nonlinear effect in $CS_2$ is significantly larger than silica, with the anisotropic molecular re-orientation being the dominant mechanism [39]. This indicates that a simulation assuming a purely isotropic polarization vector model may not accurately describe the behavior of a 2D $TM_z$ electromagnetic wave or a general 3D electromagnetic wave in some cases, potentially requiring including the anisotropic part of the polarization vector to accurately model a material like $CS_2$. A simulation of a 2D $TM_z$ electromagnetic wave propagating in $CS_2$ is shown in Fig. 4, at time $t = (\Delta t)n = 294.17$ fs as another simple proof of concept, using the extended version of the FDTD GVADE method with the polarization vector including the anisotropic part. It should be noted that the larger nonlinear effect of $CS_2$ as compared to silica can be seen in the lower required input amplitude of $4.29 \times 10^6$ A/m compared to $4.77 \times 10^7$ A/m.

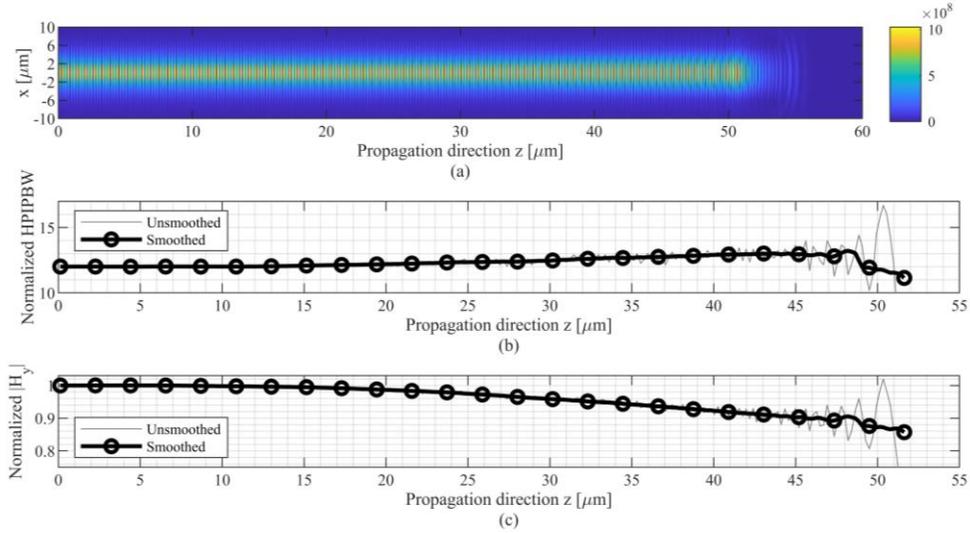

Fig. 4. Simulation results for 2D $TM_z$ wave propagation in $CS_2$ at time $t = (\Delta t)n = 294.17$ fs, using the extended FDTD GVADE method which includes the anisotropic part of the polarization vector. (a) $|\mathbf{E}|$, (b) Smoothed HPIPBW($z$) calculated relative to $\max(|H_y(z = z_{nearest})|)$, normalized to the approximate material wavelength $\lambda_d = \lambda_0/\sqrt{\varepsilon_r}$, (c) Smoothed $|H_y(z)|$ at the extremum location, normalized to the $|H_y(z = z_{nearest})|$, where $z_{nearest}$ is the nearest extremum location to $z = 0$. Note: The thinner gray curve is the "unsmoothed" curve, illustrating the "moving-mean" smoothing process used for creating the "smoothed" curves.

The original FDTD GVADE method would not be able to simulate the anisotropic aspect of the re-orientational behavior present in a material like $CS_2$ for general 3D electromagnetic waves. While for many cases, the purely isotropic polarization vector approach of the FDTD GVADE method seems to work fairly well, such as for the fused silica example above, the method presented in this paper while being more computationally expensive, may result in more accurate simulations of electromagnetic waves in isotropic media with larger anisotropic parts of the nonlinear polarization vector.

## 4. Conclusion

In this paper we have presented an extension to the FDTD GVADE method, allowing the simulation of electromagnetic wave propagation at optical frequencies in nonlinear isotropic material with a polarization vector which includes an anisotropic component along with the isotropic component previously included. While more computationally expensive, this method allows for more accurate simulation of electromagnetic wave propagation in nonlinear optical materials which have an anisotropic polarization vector component. The extended FDTD GVADE method was illustrated by simulating $TM_z$ electromagnetic wave propagation examples in fused silica and carbon disulfide.

**Acknowledgments.** Caleb Grimms thanks Junseob Kim for his contribution by creating the initial 2D isotropic polarization vector FDTD GVADE code for his master's thesis based on [1-3].

**Disclosures.** The authors declare no conflicts of interest.

**Data availability.** To the best of the authors' knowledge, the data underlying the results presented in this paper are included in this paper but may also be obtained from the authors upon reasonable request.

**Honoring.** Part of the authors' goal in writing this paper was to honor the late Dr. Allen Taflove and his contribution to the field of computational electromagnetics and the finite-difference time-domain method.

**Supplemental document.** See Supplement 1 for supporting content.

**References**

1. J. H. Greene, "Finite-difference time-domain model of resonator coupling and nonstationary nonparaxial spatial optical soliton focusing and scattering," in *Field of Electrical Engineering and Computer Science*(NORTHWESTERN UNIVERSITY, 2007), p. 95.
2. J. H. Greene, and A. Taflove, "General vector auxiliary differential equation finite-difference time-domain method for nonlinear optics," Opt Express 14, 8305-8310 (2006).
3. Z. Lubin, J. H. Greene, and A. Taflove, "GVADE FDTD Modeling of Spatial Solitons," in *Artech House Antenna Proc.*(2013), pp. 497-517, chap. 17.
4. A. Taflove, and S. C. Hagness, *Computational electrodynamics : the finite-difference time-domain method* (Artech House, 2005).
5. K. S. Yee, "Numerical Solution of Initial Boundary Value Problems Involving Maxwells Equations in Isotropic Media," Ieee T Antenn Propag Ap. 14, 302-307 (1966).
6. R. M. Joseph, and A. Taflove, "Spatial Soliton Deflection Mechanism Indicated by Fd-Td Maxwells Equations Modeling," Ieee Photonic Tech L 6, 1251-1254 (1994).
7. T. Kashiwa, and I. Fukai, "A treatment by the FD-TD method of the dispersive characteristics associated with electronic polarization," Microwave and Optical Technology Letters 3, 203-205 (1990).
8. S. Nakamura, Y. Koyamada, N. Yoshida, et al., "Finite-difference time-domain calculation with all parameters of Sellmeier's fitting equation for 12-fs laser pulse propagation in a silica fiber," Ieee Photonic Tech Lett. 14, 480-482 (2002).
9. S. Nakamura, N. Takasawa, and Y. Koyamada, "Comparison between finite-difference time-domain calculation with all parameters of Sellmeier's fitting equation and experimental results for slightly chirped 12-fs laser pulse propagation in a silica fiber," Journal of Lightwave Technology 23, 855-863 (2005).
10. M. Fujii, M. Tahara, I. Sakagami, et al., "High-order FDTD and auxiliary differential equation formulation of optical pulse propagation in 2-D Kerr and Raman nonlinear dispersive media," Ieee J Quantum Elect 40, 175-182 (2004).
11. P. M. Goorjian, and A. Taflove, "Direct Time Integration of Maxwells Equations in Nonlinear Dispersive Media for Propagation and Scattering of Femtosecond Electromagnetic Solitons," Opt Lett 17, 180-182 (1992).
12. P. M. Goorjian, A. Taflove, R. M. Joseph, et al., "Computational Modeling of Femtosecond Optical Solitons from Maxwell Equations," Ieee J Quantum Elect 28, 2416-2422 (1992).
13. R. M. Joseph, and A. Taflove, "FDTD Maxwell's equations models for nonlinear electrodynamics and optics," Ieee T Antenn Propag 45, 364-374 (1997).
14. M. Nauta, M. Okoniewski, and M. Potter, "FDTD Method on a Lebedev Grid for Anisotropic Materials," Ieee T Antenn Propag 61, 3161-3171 (2013).
15. J. J. Liu, M. Brio, and J. V. Moloney, "An overlapping Yee finite-difference time-domain method for material interfaces between anisotropic dielectrics and general dispersive or perfect electric conductor media," Int J Numer Model El 27, 22-33 (2014).
16. M. Salmasi, M. E. Potter, and M. Okoniewski, "Coupling Yee Grid to Lebedev Grid in Two Dimensions," Ieee Antennas and Wireless Propagation Letters 16, 1755-1758 (2017).
17. K. P. Prokopidis, and D. C. Zografopoulos, "Time-Domain Studies of General Dispersive Anisotropic Media by the Complex-Conjugate Pole-Residue Pairs Model," Appl Sci-Basel 11 (2021).
18. A. A. Al-Jabr, M. A. Alsunaidi, T. Khee, et al., "A Simple FDTD Algorithm for Simulating EM-Wave Propagation in General Dispersive Anisotropic Material," Ieee T Antenn Propag 61, 1321-1326 (2013).

19. H. Mosallaei, "FDTD-PLRC technique for Modeling of anisotropic-dispersive media and metamaterial devices," Ieee T Electromagn C 49, 649-660 (2007).
20. P. D. Maker, and R. W. Terhune, "Study of Optical Effects Due to an Induced Polarization Third Order in Electric Field Strength," Phys Rev 137, A801-A818 (1965).
21. P. D. Maker, R. W. Terhune, and C. M. Savage, "Intensity-Dependent Changes in Refractive Index of Liquids," Phys Rev Lett 12, 507-509 (1964).
22. C. C. Wang, "Nonlinear Susceptibility Constants and Self-Focusing of Optical Beams in Liquids," Phys Rev 152, 149-156 (1966).
23. D. H. Close, C. R. Giuliano, R. W. Hellwarth, et al., "Self-Focusing of Light of Different Polarizations," Ieee J Quantum Elect Qe 2, 553-557 (1966).
24. E. Garmire, R. Y. Chiao, and C. H. Townes, "Dynamics and Characteristics of Self-Trapping of Intense Light Beams," Phys Rev Lett 16, 347-349 (1966).
25. R. Hellwarth, J. Cherlow, and T. T. Yang, "Origin and Frequency-Dependence of Nonlinear Optical Susceptibilities of Glasses," Phys Rev B 11, 964-967 (1975).
26. R. W. Hellwarth, A. Owyoung, and N. George, "Origin of the Nonlinear Refractive Index of Liquid CCL4," Physical Review A 4, 2342-2347 (1971).
27. A. Owyoung, R. W. Hellwarth, and N. George, "Intensity-Induced Changes in Optical Polarizations in Glasses," Phys Rev B-Solid St 5, 628-633 (1972).
28. R. W. Hellwarth, "3rd-Order Optical Susceptibilities of Liquids and Solids," Prog Quant Electron 5, 1-68 (1977).
29. R. Y. Chiao, E. Garmire, and C. H. Townes, "Self-Trapping of Optical Beams," Phys Rev Lett 13, 479-482 (1964).
30. T. K. Gustafson, P. L. Kelley, R. Y. Chiao, et al., "Self-Trapping in Media with Saturation of Nonlinear Index," Appl Phys Lett 12, 165-168 (1968).
31. P. L. Kelley, "Self-Focusing of Optical Beams," Phys Rev Lett 15, 1005-1008 (1965).
32. Y. S. Kivshar, and G. P. Agrawal, *Optical solitons : from fibers to photonic crystals* (Academic Press, 2003).
33. C. A. Balanis, *Advanced engineering electromagnetics* (John Wiley & Sons, 2012).
34. G. P. Agrawal, *Nonlinear fiber optics* (Academic Press, 2019).
35. D. McMorrow, N. Thantu, V. Kleiman, et al., "Analysis of intermolecular coordinate contributions to third-order ultrafast spectroscopy of liquids in the harmonic oscillator limit," J Phys Chem A 105, 7960-7972 (2001).
36. K. J. Blow, and D. Wood, "Theoretical Description of Transient Stimulated Raman-Scattering in Optical Fibers," Ieee J Quantum Elect 25, 2665-2673 (1989).
37. D. McMorrow, N. Thantu, J. S. Melinger, et al., "Probing the microscopic molecular environment in liquids: Intermolecular dynamics of CS2 in alkane solvents," J Phys Chem-Us 100, 10389-10399 (1996).
38. Q. Lin, and G. P. Agrawal, "Raman response function for silica fibers," Opt Lett 31, 3086-3088 (2006).
39. M. Reichert, H. H. Hu, M. R. Ferdinandus, et al., "Temporal, spectral, and polarization dependence of the nonlinear optical response of carbon disulfide," Optica 1, 436-445 (2014).
40. M. Reichert, H. H. Hu, M. R. Ferdinandus, et al., "Temporal, spectral, and polarization dependence of the nonlinear optical response of carbon disulfide: erratum (vol 1, pg 436, 2014)," Optica 3, 657-658 (2016).
41. M. Chemnitz, M. Gebhardt, C. Gaida, et al., "Hybrid soliton dynamics in liquid-core fibres," Nat Commun 8 (2017).
42. D. Kincaid, and E. W. Cheney, *Numerical analysis : mathematics of scientific computing* (Brooks/Cole, 2002).
43. Y. Liu, "Fourier analysis of numerical algorithms for the Maxwell equations," J Comput Phys 124, 396-416 (1996).
44. C. J. Grimms, and R. D. Nevels, "Rotationally symmetric transverse magnetic vector wave propagation for nonlinear optics," Opt. Express 32, 17853-17868 (2024).
45. R. H. Stolen, J. P. Gordon, W. J. Tomlinson, et al., "Raman Response Function of Silica-Core Fibers," J Opt Soc Am B 6, 1159-1166 (1989).
46. M. Reichert, H. H. Hu, M. R. Ferdinandus, et al., "Temporal, spectral, and polarization dependence of the nonlinear optical response of carbon disulfide: supplementary material," Optica 1 (2014).

# EXTENDING THE FDTD GVADE METHOD NONLINEAR POLARIZATION VECTOR TO INCLUDE ANISOTROPY: SUPPLEMENTAL DOCUMENT

To further understand the rank 4 susceptibility tensor composed of 81 terms and the third order polarization vector equations, index notation is introduced.

## Section 1: The polarization vector and rank 4 susceptibility tensor using index notation

The summations are explicitly shown in the following equations for clarity. Re-writing equation (5) using index notation ($i, j, k, l = x, y, z$) [1],

$$P_i^{NL}(\mathbf{r},t) = \varepsilon_0 \sum_j \sum_k \sum_l \int_{-\infty}^{\infty}\int_{-\infty}^{\infty}\int_{-\infty}^{\infty} \left[\chi_{ijkl}^{(3)}(t-t_1, t-t_2, t-t_3) \; E_j(\mathbf{r},t_1)E_k(\mathbf{r},t_2)E_l(\mathbf{r},t_3)\right] dt_1 dt_2 dt_3 \;. \tag{S1}$$

Next, re-writing equations (7) and (8) following [1], where $P_i^{NL}(\mathbf{r},t) = P_{el,i}^{NL}(\mathbf{r},t) + P_{nu,i}^{NL}(\mathbf{r},t)$,

$$P_{el,i}^{NL}(\mathbf{r},t) = \varepsilon_0 \sum_j \sum_k \sum_l \chi_{el,ijkl}^{(3)} E_j(\mathbf{r},t)E_k(\mathbf{r},t)E_l(\mathbf{r},t) \;, \tag{S2}$$

$$P_{nu,i}^{NL}(\mathbf{r},t) = \varepsilon_0 \sum_j \sum_k \sum_l E_j(\mathbf{r},t) \int_{-\infty}^{\infty} \left[\chi_{nu,ijkl}^{(3)}(t-t')E_k(\mathbf{r},t')E_l(\mathbf{r},t')\right] dt' \;. \tag{S3}$$

The general rank 4 isotropic susceptibility tensor used for isotropic media is shown for the "nuclear" part [1-3], where $\delta_{ij}=1$ when $i=j$ and $\delta_{ij}=0$ when $i \neq j$:

$$\chi_{nu,ijkl}^{(3)}(t) = \chi_{nu,a}^{(3)}(t)\delta_{ij}\delta_{kl} + \frac{1}{2}\chi_{nu,b}^{(3)}(t)\left(\delta_{ik}\delta_{jl} + \delta_{il}\delta_{jk}\right) \;. \tag{S4}$$

The term $\chi_{nu,a}^{(3)}(t)(\delta_{ij}\delta_{kl})$, after being plugged into equation (S3), produces the isotropic part of the nonlinear polarization vector,

$$P_{nu,a,i}^{NL}(\mathbf{r},t) = \varepsilon_0 \sum_j \sum_k \sum_l E_j(\mathbf{r},t) \int_{-\infty}^{\infty} \left[\chi_{nu,a}^{(3)}(t-t')\delta_{ij}\delta_{kl}E_k(\mathbf{r},t')E_l(\mathbf{r},t')\right] dt' \;, \tag{S5}$$

and the $(1/2)\chi_{nu,b}^{(3)}(t)(\delta_{ik}\delta_{jl}+\delta_{il}\delta_{jk})$ term, after being plugged into equation (S3), produces the anisotropic part of the nonlinear polarization vector,

$$P_{nu,b,i}^{NL}(\mathbf{r},t) = \varepsilon_0 \sum_j \sum_k \sum_l E_j(\mathbf{r},t) \int_{-\infty}^{\infty} \left[\frac{1}{2}\chi_{nu,b}^{(3)}(t-t')\left(\delta_{ik}\delta_{jl} + \delta_{il}\delta_{jk}\right) E_k(\mathbf{r},t')E_l(\mathbf{r},t')\right] dt' \;. \tag{S6}$$

The electronic contribution, being much faster than the nuclear response, is [4] approximated as being instantaneous using the Dirac delta function or impulse response, $g_{el}(t) = \delta(t)$ [1]:

$$\chi_{el,ijkl}^{(3)}(t) = \chi_{el}^{(3)}\delta(t)\frac{1}{3}\left(\delta_{ij}\delta_{kl} + \delta_{ik}\delta_{jl} + \delta_{il}\delta_{jk}\right) \;. \tag{S7}$$

When equation (S4) is plugged into equation (S3) it results in equation (9), and when equation (S7) is plugged into equation (S2) it results in equation (7). The electronic contribution and the nuclear contributions can also be written together as,

$$\chi^{(3)}_{ijkl}(t) = \frac{1}{3}\chi^{(3)}_{el}\delta(t)\left(\delta_{ij}\delta_{kl}+\delta_{ik}\delta_{jl}+\delta_{il}\delta_{jk}\right) + \chi^{(3)}_{nu,a}(t)\delta_{ij}\delta_{kl} + \frac{1}{2}\chi^{(3)}_{nu,b}(t)\left(\delta_{ik}\delta_{jl}+\delta_{il}\delta_{jk}\right), \tag{S8}$$

where $\chi^{(3)}(t)$ from equation (5) is related to the index notation $\chi^{(3)}_{ijkl}(t)$ using the summation of the unit vector dyadic products, dyads, $\hat{a}_i\hat{a}_j$ and $\hat{a}_k\hat{a}_l$ with $\chi^{(3)}_{ijkl}(t)$,

$$\chi^{(3)}(t) = \sum_i\sum_j\sum_k\sum_l \hat{a}_i\hat{a}_j\,\chi^{(3)}_{ijkl}(t)\,\hat{a}_k\hat{a}_l\ . \tag{S9}$$

A constant scalar value $\chi^{(3)}_0$ can be pulled out of $\chi^{(3)}_{ijkl}(t)$, leading to,

$$\chi^{(3)}_{ijkl}(t) = \chi^{(3)}_0 g_{ijkl}(t) = \chi^{(3)}_0 \left[\frac{1-f_R}{3}\delta(t)\left(\delta_{ij}\delta_{kl}+\delta_{ik}\delta_{jl}+\delta_{il}\delta_{jk}\right) + f_{nu,a}g_{nu,a}(t)\delta_{ij}\delta_{kl} + \frac{1}{2}f_{nu,b}g_{nu,b}(t)\left(\delta_{ik}\delta_{jl}+\delta_{il}\delta_{jk}\right)\right], \tag{S10}$$

where $g_{nu,a}(t)$ and $g_{nu,b}(t)$ are the normalized isotropic and anisotropic parts of the normalized nuclear response function $g_{ijkl}(t)$ according to [5], normalized so that $\int_0^\infty g_{nu,a}(t)dt=1$ and $\int_0^\infty g_{nu,b}(t)dt=1$, with relative strengths $f_{nu,a}$ and $f_{nu,b}$ defined as $f_R = f_{nu,a}+f_{nu,b}$ and $f_{el}=\left(1-f_R\right)$. The normalization of $g_{ijkl}(t)$ can be seen by looking at $i=j=k=l=x$, resulting in $g_{xxxx}(t) = [(1-f_R)\delta(t) + f_{nu,a}g_{nu,a}(t) + f_{nu,b}g_{nu,b}(t)]\delta_{xx}\delta_{xx}$, where $\int_0^\infty g_{xxxx}(t)dt=1$, since $\delta_{xx}\delta_{xx}=1$ [6]. This example shows that the 1/3 and the 1/2 factors in equation (S10), are necessary for normalization due to the tensor nature of the polarization vector. For example, the summation of the electronic $(\delta_{ij}\delta_{kl}+\delta_{ik}\delta_{jl}+\delta_{il}\delta_{jk})$ term results in $3(\varepsilon_0\chi^{(3)}_{el}\mathbf{E}(\mathbf{r},t)|\mathbf{E}(\mathbf{r},t)|^2)$ requiring the 1/3 factor to normalize the term. In a similar manner the summation of the nuclear "anisotropic" $(\delta_{ik}\delta_{jl}+\delta_{il}\delta_{jk})$ term results in two times the anisotropic convolution integral, requiring the 1/2 factor to normalize it. This definition of $\chi^{(3)}_{ijkl}(t)$ can be plugged into,

$$P^{NL}_i(\mathbf{r},t) = \varepsilon_0 \sum_j\sum_k\sum_l E_j(\mathbf{r},t)\int_{-\infty}^{\infty}\left[\chi^{(3)}_{ijkl}(t-t')E_k(\mathbf{r},t')E_l(\mathbf{r},t')\right]dt', \tag{S11}$$

resulting in equations (6), (7) and (8). This substitution allows equations (S2), (S5) and (S6) to be re-written in terms of $f_{el}$, $f_{nu,a}$ and $f_{nu,b}$ as,

$$P^{NL}_{el,i}(\mathbf{r},t) = \varepsilon_0\chi^{(3)}_0 \sum_j\sum_k\sum_l \frac{1}{3}f_{el}\delta(t)\left(\delta_{ij}\delta_{kl}+\delta_{ik}\delta_{jl}+\delta_{il}\delta_{jk}\right)E_j(\mathbf{r},t)E_k(\mathbf{r},t)E_l(\mathbf{r}, \tag{S12}$$

$$P^{NL}_{nu,a,i}(\mathbf{r},t) = \varepsilon_0\chi^{(3)}_0 \sum_j\sum_k\sum_l E_j(\mathbf{r},t)\int_{-\infty}^{\infty}\left[f_{nu,a}\,g_{nu,a}(t-t')\delta_{ij}\delta_{kl}\,E_k(\mathbf{r},t')E_l(\mathbf{r},t')\right]dt', \tag{S13}$$

$$P^{NL}_{nu,b,i}(\mathbf{r},t) = \varepsilon_0\chi^{(3)}_0 \sum_j\sum_k\sum_l E_j(\mathbf{r},t)\int_{-\infty}^{\infty}\left[\frac{1}{2}f_{nu,b}\,g_{nu,b}(t-t')\left(\delta_{ik}\delta_{jl}+\delta_{il}\delta_{jk}\right)E_k(\mathbf{r},t')E_l(\mathbf{r},t')\right] \tag{S14}$$

which are equivalent to equations (7) and (9).

## Section 2: Silica rank 4 susceptibility tensor

The normalized nonlinear rank 4 material response tensor for silica written in a slightly modified form from [6] is,

$$g^{(3)}_{ijkl}(t) = \left[\frac{f_{el}}{3}\delta(t)\left(\delta_{ij}\delta_{kl} + \delta_{ik}\delta_{jl} + \delta_{il}\delta_{jk}\right) + \right. \tag{S15}$$
$$\left. f_a g_{nu,a}(t)\delta_{ij}\delta_{kl} + \frac{1}{2}\left[f_b g_b(t) + f_c g_{nu,a}(t)\right]\left(\delta_{ik}\delta_{jl} + \delta_{il}\delta_{jk}\right)\right].$$

For comparison, the normalized isotropic nonlinear polarization vector response function is

$$g^{(3)}_{ijkl}(t) = \left[\frac{(1-f_{R,\text{Isotropic}})}{3}\delta(t)\left(\delta_{ij}\delta_{kl}+\delta_{ik}\delta_{jl}+\delta_{il}\delta_{jk}\right) + f_{R,\text{Isotropic}}g_{nu,a}(t)\delta_{ij}\delta_{kl}\right], \tag{S16}$$

where $f_{R,\text{Isotropic}}$ is slightly different then $f_R$ as described by [6, 7].

## Section 3: Carbon disulfide rank 4 susceptibility tensor

The nonlinear rank 4 tensor for $CS_2$ written in a slightly modified form from [6] is,

$$g^{(3)}_{ijkl}(t) = \left[\frac{f_{el}}{3}\delta(t)\left(\delta_{ij}\delta_{kl} + \delta_{ik}\delta_{jl} + \delta_{il}\delta_{jk}\right) + \right. \tag{S17}$$
$$\left. f_c g_c(t)\delta_{ij}\delta_{kl} + \frac{1}{2}\left[f_d g_d(t) + f_v g_v(t)\right]\left(\delta_{ik}\delta_{jl} + \delta_{il}\delta_{jk}\right)\right].$$

where $g_c(t)$, $g_v(t)$ and $g_d(t)$ correspond to the (overdamped) "collision", (underdamped) "libration" resulting from molecular re-orientation, and (overdamped) "diffusive molecular re-orientation" nuclear response functions for $CS_2$ [8, 9].

The electronic and nuclear parts of the susceptibility can be calculated as $\chi^{(3)}_{el} = (4/3)c\,\varepsilon_0\varepsilon_r n_{2,el}$ and $\chi^{(3)}_{nu}(t) = 2\,c\,\varepsilon_0\varepsilon_r R(t)$ where $R(t) = \sum_m n_{2,m}g_m(t)$ as described in [8, 10]. To determine $\chi^{(3)}_0$, $f_{el}$, $f_c$, $f_l$ and $f_d$, the special case where $i=j=k=l=x$ was considered leading to,

$$\chi^{(3)}_{xxxx}(t) = (4/3)c\,\varepsilon_0\varepsilon_r n_{2,el}g_{el}(t) + 2\,c\,\varepsilon_0\varepsilon_r\left[n_{2,c}g_c(t) + n_{2,l}g_l(t) + n_{2,d}g_d(t)\right]. \tag{S18}$$

The equation can be re-written pulling out a constant $\chi^{(3)}_0$ as $\chi^{(3)}_{xxxx}(t) = \chi^{(3)}_0 g_{xxxx}(t)$,

$$\chi^{(3)}_{xxxx}(t) = \chi^{(3)}_0\left[f_{el}g_{el}(t) + f_c g_c(t) + f_l g_l(t) + f_d g_d(t)\right], \tag{S19}$$

where $f_{el} = (2/3)n_{2,el}/n_{2,\text{sum}}$, $f_c = n_{2,c}/n_{2,\text{sum}}$, $f_l = n_{2,l}/n_{2,\text{sum}}$, $f_d = n_{2,d}/n_{2,\text{sum}}$, $\chi^{(3)}_0 = 2\,c\,\varepsilon_0\varepsilon_r n_{2,\text{sum}}$, and $n_{2,\text{sum}} = (2/3)n_{2,el} + n_{2,c} + n_{2,l} + n_{2,d}$. The resulting values were $\chi^{(3)}_0 = 3.81677\times10^{-20}\,\text{m}^2/\text{V}^2$, $f_{el} = 0.0362$, $f_c = 0.0362$, $f_l = 0.2754$ and $f_d = 0.65217$.

### References

1. R. W. Hellwarth, "3rd-Order Optical Susceptibilities of Liquids and Solids," Prog Quant Electron 5, 1-68 (1977).
2. P. M. Morse, and H. Feshbach, *Methods of theoretical physics* (McGraw-Hill, 1953).
3. R. P. Feynman, R. B. Leighton, and M. L. Sands, *The Feynman lectures on physics* (Basic Books, 2011).
4. R. W. Boyd, *Nonlinear optics* (Academic Press is an imprint of Elsevier, 2019).
5. G. P. Agrawal, *Nonlinear fiber optics* (Elsevier / Academic Press, 2007).
6. Q. Lin, and G. P. Agrawal, "Raman response function for silica fibers," Opt Lett 31, 3086-3088 (2006).
7. G. P. Agrawal, *Nonlinear fiber optics* (Academic Press, 2019).
8. M. Reichert, H. H. Hu, M. R. Ferdinandus, et al., "Temporal, spectral, and polarization dependence of the nonlinear optical response of carbon disulfide," Optica 1, 436-445 (2014).
9. M. Reichert, H. H. Hu, M. R. Ferdinandus, et al., "Temporal, spectral, and polarization dependence of the nonlinear optical response of carbon disulfide: erratum (vol 1, pg 436, 2014)," Optica 3, 657-658 (2016).

10. M. Reichert, H. H. Hu, M. R. Ferdinandus, et al., "Temporal, spectral, and polarization dependence of the nonlinear optical response of carbon disulfide: supplementary material," Optica 1 (2014).